%% file: dac2019.tex
\documentclass[sigconf]{acmart}

\usepackage[]{algpseudocode}
\usepackage[]{algorithm}
\usepackage{multirow}
\usepackage{booktabs} 
\usepackage{arydshln}
\usepackage{relsize}

\copyrightyear{2019}
\acmYear{2019}
\setcopyright{acmlicensed}
\acmConference[DAC '19]{The 56th Annual Design Automation Conference 2019}{June 2--6, 2019}{Las Vegas, NV, USA}
\acmBooktitle{The 56th Annual Design Automation Conference 2019 (DAC '19), June 2--6, 2019, Las Vegas, NV, USA}
\acmPrice{15.00}
\acmDOI{10.1145/3316781.3317781}
\acmISBN{978-1-4503-6725-7/19/06}

\newcommand{\affFIT}{$^{1}$}
\newcommand{\affTU}{$^{2}$}

\begin{document}
\title{autoAx: An Automatic Design Space Exploration and Circuit Building Methodology utilizing Libraries of Approximate Components}
\renewcommand{\shorttitle}{autoAx: An Automatic Design Space Exploration and Circuit Building Methodology}

\author{Vojtech Mrazek$^{1,2}$} 
\affiliation{}
\email{mrazek@fit.vutbr.cz}

\author{Muhammad Abdullah Hanif\affTU}
\affiliation{}
\email{muhammad.hanif@tuwien.ac.at}

\author{Zdenek Vasicek\affFIT}
\affiliation{}
\email{vasicek@fit.vutbr.cz}

\author{Lukas Sekanina\affFIT}
\affiliation{}
\email{sekanina@fit.vutbr.cz}

\author{Muhammad Shafique\affTU}
\affiliation{}
\email{muhammad.shafique@tuwien.ac.at}

\newcommand\sharedaffiliation{  
\affFIT Faculty of Information Technology, IT4Innovations Centre of Excellence, Brno University of Technology, Czech Republic\\
\affTU Institute of Computer Engineering, Vienna University of Technology (TU Wien), Austria
}

\renewcommand{\shortauthors}{V. Mrazek, M. A. Hanif, Z. Vasicek, L. Sekanina and M. Shafique}

\begin{abstract}
Approximate computing is an emerging paradigm for developing highly energy-efficient computing systems such as various accelerators. In the literature, many libraries of elementary approximate circuits have already been proposed to simplify the design process of approximate accelerators. Because these libraries contain from tens to thousands of approximate implementations for a single arithmetic operation it is intractable to find an optimal combination of approximate circuits in the library even for an application consisting of a few operations. An open problem is ``how to effectively combine circuits from these libraries to construct complex approximate accelerators''. This paper proposes a novel methodology for \textit{searching, selecting} and \textit{combining} the most suitable approximate circuits from a set of available libraries to generate an approximate accelerator for a given application. To enable fast design space generation and exploration, the methodology utilizes machine learning techniques to create  computational models estimating the overall quality of processing and hardware cost without performing full synthesis at the accelerator  level. Using the methodology, we construct hundreds of approximate accelerators (for a Sobel edge detector) showing different but relevant tradeoffs between the quality of processing and hardware cost and identify a corresponding Pareto-frontier. Furthermore, when searching for  approximate implementations of a generic Gaussian filter consisting of 17 arithmetic operations, the proposed approach allows us to identify approximately $10^3$ highly relevant implementations from $10^{23}$ possible solutions in a few hours, while the exhaustive search would take four months on a high-end processor.
\end{abstract}

%
%



\maketitle

\input{dac2019-body.tex}

\bibliographystyle{ACM-Reference-Format}
\bibliography{dac2019}

\end{document}

%% file: dac2019-body.tex
\section{Introduction}

Approximate computing is an emerging paradigm that allows to develop highly energy-efficient computing systems such as various hardware accelerators for image filtering, video processing and data mining. It capitalizes inherent error resilience of many applications to trade Quality of Result (QoR) with energy efficiency.  At the circuit level, functional approximation is achieved by employing approximate implementations for carefully selected operations of the accelerator. Literature contains a good body of works dealing with automated design methods for approximate circuits, e.g., CGP~\cite{vasicek:TEC}, SALSA~\cite{Salsa}, SASIMI~\cite{sasimi}. Majority of these works focus on elementary approximate circuits such as approximate adders and multipliers because they are building blocks of many applications.  Approximate implementations of arithmetic circuits can also be downloaded (at the level of synthesized netlist or C code) from open source libraries such as~\cite{mrazek:date16lib} or created using quality-configurable approximate structures (such as QuAd adders~\cite{hanif:quad}, GeAR adders~\cite{Shafique:gear2015}, structured multipliers~\cite{Rehman:2016} or Broken-array multipliers (BAM)~\cite{jiang,Mahdiani:2010}). All the approximate circuits available in these libraries are fully characterized in terms of electrical properties and various error metrics. 

Because these libraries contain from tens to thousands of approximate implementations for each arithmetic operation, the user is provided with a broad set of implementation options to reach the best possible tradeoff between QoR and energy (or other hardware parameters) at the accelerator level. However, \textbf{it is intractable to find an optimal combination of approximate circuits} even for an accelerator consisting of a few operations. The problem addressed in this paper is to \textit{identify the most suitable replacement of arithmetic operations of target accelerator with approximate circuits available in the library}. As it is a multi-objective optimization problem, there is no single optimal solution, rather multiple ones typically exist. We are primarily interested in approximate circuits belonging to the \emph{Pareto frontier} that contains the so-called non-dominated solutions. Consider two objectives to be minimized, for example, the mean error and energy. Circuit C1 (Pareto) \emph{dominates} another circuit C2 if: 1) C1 is no worse than C2 in all objectives and 2) C1 is strictly better than C2 in at least one objective.

This problem resembles the binding step of the high-level synthesis (HLS) whose objective is to (i) map elementary operations of the algorithm to specific instances of components that are available in the component library, and (ii) optimize hardware parameters such as latency, area and power consumption. In the context of approximate circuits, the principal difference and \textbf{difficulty lies in the QoR evaluation} at the accelerator level. Except some very specific cases (e.g.~\cite{Mazahir:tc17mult,Mazahir:2017adders}), it is in general unknown how the errors propagate if two or  more approximate circuits are connected in a more complex circuit. A common approach is to estimate the resulting error using either analytic or statistical techniques, but it usually is a very unreliable approach as seen in~\cite{Li:2015}. If the problem is simplified in such a way that the only approximation technique is truncation then an optimal number of bits to be approximated can be determined~\cite{Sengupta:dac17}.
 
\textbf{Proposed Methodology:} In this paper, our objective is to identify the most suitable replacement of arithmetic operations (of the  original accelerator) with approximate circuits. It is assumed that approximate circuits available in a library are fully characterized (in terms of error and hardware parameters), but nothing is assumed about their internal structure (i.e., an arbitrary approximation technique can be used to build the elementary approximate circuit, not only truncation). As a huge number of candidate replacements exist, the key idea is to eliminate as many clearly sub-optimal solutions as possible without performing precise evaluation of QoR and time-consuming circuit synthesis at the accelerator level. In order to estimate QoR, we propose to build a computational model  using the error metrics (which are pre-calculated for each approximate circuit in the library) and machine learning techniques. The error model is then used to estimate QoR of candidate designs during the design space exploration process. Similarly, another computational model is constructed and applied to estimate hardware parameters in the design space exploration process. A similar approach has already been applied for common circuits on FPGAs~\cite{dai:2018}. In the context of approximate computing, machine learning techniques were applied to estimate QoR in the design of approximate accelerators in which the approximations are based on using multiple voltage islands~\cite{zervakis:2018}.  In our methodology, due to the enormous number of possible candidate solutions, the resulting Pareto frontier is identified using a hill climbing algorithm which works with estimated QoR and estimated hardware parameters.

\textbf{Novel Contributions:} In this paper, we propose  a novel methodology for \textit{searching, selecting} and \textit{combining} the most suitable approximate circuits from a set of available libraries to generate an approximate accelerator for a given application. \textbf{To address the aforementioned scientific challenges, in this paper, we make the following key contributions.} 
(i) A new QoR estimation technique is developed, which is based on computational models constructed using machine learning methods. This technique works with arbitrary approximate circuits, i.e., not only with those created by truncation or other well-understood methods.
(ii) A new heuristic Pareto frontier construction algorithm, based on proposed estimation techniques, is presented and evaluated.
(iii) The proposed methodology is evaluated using three case studies (Sobel edge detector, Gaussian filter with fixed coefficients and Generic Gaussian filter) in which approximate accelerators showing high-quality tradeoffs between QoR and hardware parameters are generated in a fully automated way using a library containing thousands of approximate circuits. The proposed method significantly reduces the number of design alternatives that have to be considered and evaluated.

\vspace{-0.6em}
\section{Proposed methodology}
\vspace{-0.3em}
\subsection{Overview}
The methodology requires the following input from the user: a hardware description of the chosen accelerator, corresponding software model and training (benchmark) data. 
Hierarchical hardware as well as software models are expected in order to be able to replace relevant operations with their approximate versions, and to evaluate how this change affects the QoR. Approximate circuits are taken from a library, in which each of them is fully characterized and many approximate implementations exist for each operation. 

Let the accelerator contain $n$ operations that can be implemented using some approximate circuits for the library. By \emph{configuration} we mean a particular assignment of approximate circuits from the library to $n$ operations of the accelerator. The goal of the methodology is to find a Pareto set of \textit{configurations} where the design objectives to be optimized are QoR (e.g., SSIM, PSNR etc.) and hardware cost (e.g., area, delay, power or energy).


\begin{figure}[tb]
    \centering
    \includegraphics[width=0.85\columnwidth]{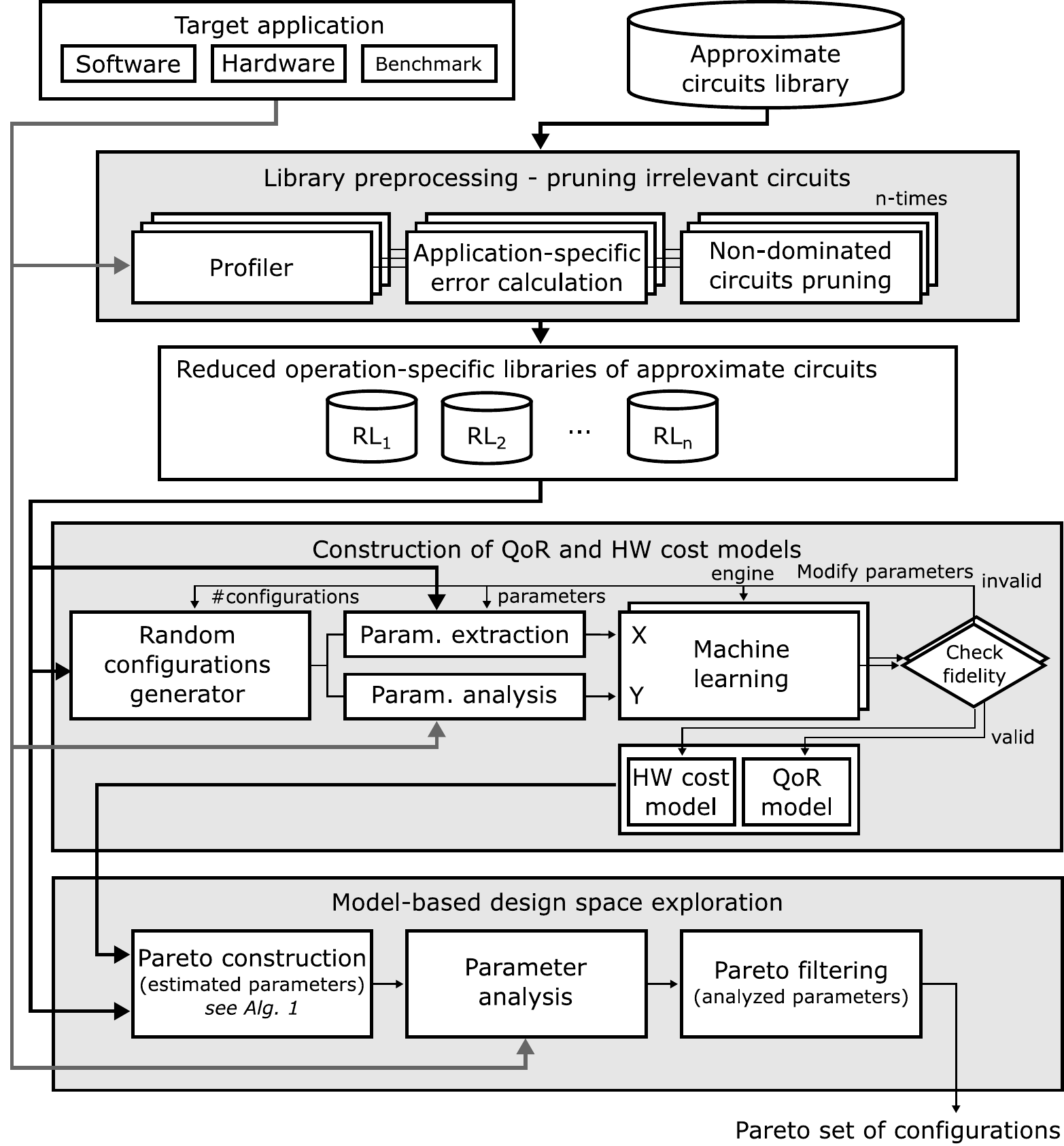}\vspace{-1em}
    \caption{Overview of the proposed \textit{autoAx} methodology.}\label{fig:arch}
    \vspace{-7mm}
\end{figure}

The whole process consists of three steps as illustrated in Figure~\ref{fig:arch}.\\
\textit{Step 1:} The library of the approximate circuits is pre-processed in such a way that clearly irrelevant circuits are removed. The irrelevant circuits are identified on the basis of their quality (measured w.r.t. a particular application) and hardware cost.\\
\textit{Step 2:} Computational models enabling to estimate QoR and hardware cost are constructed by means of some machine learning algorithm. 
A small (randomly selected) subset of possible configurations  is used for learning of the computational models.\\
\textit{Step 3:} The Pareto frontier reflecting QoR and HW cost is constructed. To quickly remove as many low-quality solutions as possible, the construction algorithm employs the values estimated by the proposed models. The final Pareto front is then constructed using precisely computed QoR and hardware parameters by means of simulation and synthesis.

\subsection{Library pre-processing}
For each operation of the accelerator, a suitable subset of approximate circuits is separately identified in the library by means of benchmark data. For example, if $k$-th operation of the accelerator is 8-bit addition then the objective of this step is to identify approximate 8 bit adders that form the Pareto front with respect to a suitable error metric (score) and hardware cost. 
We propose to base the selection on probability mass function (PMF) of the given operation which can be easily determined by simulation of the accelerator on benchmark data.


This process can be formalized as follows. Let $I$ denote a set of all possible combination of values from the benchmark data set that can occur on the input of $k$-th operation $M(x_1, x_2, \dots)$, $x \in I$, $k=1 \dots n$. Then, $D_k: I \rightarrow \mathbb{R}$ denoting the PMF of this operation is defined as 
$D_k(i_1,i_2,\dots) =  Pr(x_1 = i_1 \wedge x_2 = i_2 \wedge \dots)$.
This function is used to determine a score (weighted mean error distance) of an approximate circuit $\widetilde{M}$ implementing $k$-th operation as follows:
$
\mathrm{WMED_{k}(\widetilde{M})} = \sum_{\forall i \in I} D_k({i}) \cdot | M(i) - \widetilde{M}(i)|
$
For each operation of the accelerator, this score is then used together with hardware cost  to identify only those approximate circuits (i.e., 8-bit adders in our example) that are lying on a Pareto frontier. 

\subsection{Models construction}
Since the full synthesis and simulation are typically very time consuming processes, it is intractable to use them to perform the analysis of hardware cost and QoR for every possible configuration of the accelerator. To address this issue, we propose to construct two independent computational models, one for estimating QoR and a second for estimating hardware parameters. The estimation is based on the parameters of approximate circuits belonging to one selected configuration.

The models are constructed independently using a suitable supervised \textit{machine learning algorithm}. The learning process is based on providing example input--output pairs. In our case, each input--output pair corresponds with a particular configuration. 
One input is represented by a vector, which contains a subset of hardware or quality parameters of each approximate circuit realizing one of operations as defined by the configuration. The output is a single value of QoR or hardware cost that is obtained by simulation and synthesis of the concrete accelerator with  the given configuration. For learning, we have to generate a training set typically containing from hundreds or thousands of configurations.

The goal of this step is to obtain high-quality models. A set of configurations different from the training set is used to determine the quality of the model and avoid \textit{overfitting}\footnote{the estimated values correspond too closely or exactly to training output values, and the model may, therefore, therefore fail in fitting additional data}. Typically, the accuracy is optimized by the machine learning algorithms. However, as the models are used for determining a relation between two different configurations, it is not necessary to focus on the accuracy. We propose to consider \emph{fidelity} as the optimization criterion and maximize the fidelity of the model.
The fidelity tells us how often the estimated values are in the same relation ($<,=$ or $>$) as the real values for each pair of configurations. 
If the fidelity of the constructed model is insufficient, we have to tune parameters of the chosen learning algorithm or select a different learning engine. 

\subsection{Model-based design space exploration}
In this step, Pareto frontier containing those configurations that show the best tradeoffs  between QoR and hardware cost is constructed. In order to avoid time-consuming simulation and synthesis, the construction is divided into two stages. In the first stage, the computational models that we have developed in the previous step are used to build a pseudo Pareto set of potentially good configurations. 
In the second stage, based on the configurations forming the pseudo Pareto set, a set of approximate accelerators is determined, fully synthesized
and analyzed by means of a simulator and benchmark data. A real QoR and real hardware cost is assigned to each configuration. Finally, these real values are used to construct the final Pareto set. 


\newlength{\textfloatsepsave} \setlength{\textfloatsepsave}{\textfloatsep} 
\setlength{\textfloatsep}{5pt}
\begin{algorithm}[t]
\caption{Pareto set construction}\label{hillclimb}\smaller%
\input{algorithmv2.tex}

\end{algorithm}

Although the first step reduced the number of possible configurations, the number of combinations may still be
enormous especially for complex problems consisting of tens of operations. Therefore, we proposed an iterative heuristic algorithm (Algorithm \ref{hillclimb}) to construct the pseudo Pareto set.
The algorithm is a variant of stochastic hill climbing which starts with a random configuration (denoted as $Parent$), selects a neighbor at random (denoted as $C$), and decides whether to move to that neighbor or to examine another. The neighbor configuration is derived from $Parent$ by modifying a randomly chosen item of the configuration (i.e., another circuit is picked from the library for a randomly chosen operation). 
The quality and hardware cost parameters of $C$ ($e_{QoR}$ and $e_{HW}$) are estimated by means of appropriate estimation models. If the estimated values dominate those already present in  Pareto set $P$, configuration $C$ is inserted to the set, the set is updated (operation \textsc{ParetoInsert}) and the candidate is used as the $Parent$ in the next iteration. In order to avoid getting stuck in a local optimum, restarts are used. If the $Parent$ remains unchanged for $k$ successive iterations, the $Parent$ is replaced by a randomly chosen configuration from $P$.
The quality of the resulting Pareto set depends on the fidelity of the estimation models and on the number of allowed iterations. The higher fidelity, the better results. The number of iterations depends on the chosen termination condition. It can be determined by the size of $P$, execution time, or the maximum allowed number of iterations.

\section{Experimental setup}
The proposed methodology is evaluated on three accelerators of different complexity that are typically used as benchmarks in the area of image processing. In particular, Sobel edge detector (Sobel ED), Gaussian filter with fixed coefficients (Fixed GF) and Generic Gaussian filter (Generic GF) working on 3x3 filter kernel were chosen. While the approximation of the first problem is solvable by an exhaustive enumeration of all possible configurations, the Generic GF consists of 17 operations and represents a non-trivial problem. The particular instances of the operations the chosen problems consist of are reported in Table~\ref{tab:circs}. In all cases, 8-bit gray-scale images are considered at the input. The problems were described in Verilog HDL which is used for synthesis (HW model) and in C++ (SW model) which is used for QoR analysis. The images consisting of $384\times256$ pixels from Berkeley Segmentation Dataset\footnote{\url{https://www2.eecs.berkeley.edu/Research/Projects/CS/vision/bsds/}} are used as benchmark data. 
To evaluate QoR, i.e., to determine the difference between the output of approximate and accurate implementations, we chose a commonly used measure known as the \textit{structural similarity} index (SSIM). 
To determine the hardware cost, Synopsys Design Compiler targeting 45~nm ASIC technology was employed as a synthesis tool. The total \textit{area on the chip} was considered in this study as a cost parameter. 

\setlength{\textfloatsep}{\textfloatsepsave}

\begin{table}[t]\vspace{-5pt}
    \centering%
    \caption{The number of operations in target accelerators}\vspace{-1em}\smaller%
    {\setlength\tabcolsep{2pt}\begin{tabular}{p{2cm} | c c c | c c | c | c}\toprule
 & \multicolumn{3}{c|}{\bf Adder} & \multicolumn{2}{c|}{\bf Subtractor} & \bf Multiplier & \bf Total \\ 
\bf Problem & 8-bit & 9-bit & 16-bit & 10-bit & 16-bit & 8-bit & \bf \#\\\midrule
Sobel ED & 2 & 2 & -- & 1 & -- & --  & 5 \\
Fixed GF & 4 & 2 & 4  & -- & 1 & --  & 11 \\
Generic GF & -- & -- & 8  & -- & -- & 9  & 17 \\\bottomrule
   \end{tabular}}

    \label{tab:circs}
\end{table}

The approximate circuits implementing each of six operations shown in Table~\ref{tab:circs} are 
obtained from an extended version of EvoApprox~\cite{mrazek:date16lib} library. In addition to that, QuAd~\cite{hanif:quad} adders and BAM~\cite{Mahdiani:2010} multipliers are utilized. 
The total number of various circuits that are available in our initial library is shown in Table~\ref{tab:libs}.

\begin{table}[htb]\vspace{-5pt}
    \centering%
    \caption{Approximate circuits included in the library}\vspace{-1em}\smaller%
    \setlength\tabcolsep{2pt}\begin{tabular}{c | c c c | c c | c }\toprule
     & \multicolumn{3}{c|}{\bf Adder} & \multicolumn{2}{c|}{\bf Subtractor} & \bf Multiplier \\
    \bf instance & 8-bit & 9-bit & 16-bit & 10-bit & 16-bit & 8-bit  \\\midrule
    \bf \# implementations & 6979 & 332 & 884 & 365 & 460 & 29911\\\bottomrule
    \end{tabular}
    \vspace{-1em}
    \label{tab:libs}
\end{table}

\begin{figure}[th]
    \centering
    \includegraphics[width=0.9\columnwidth]{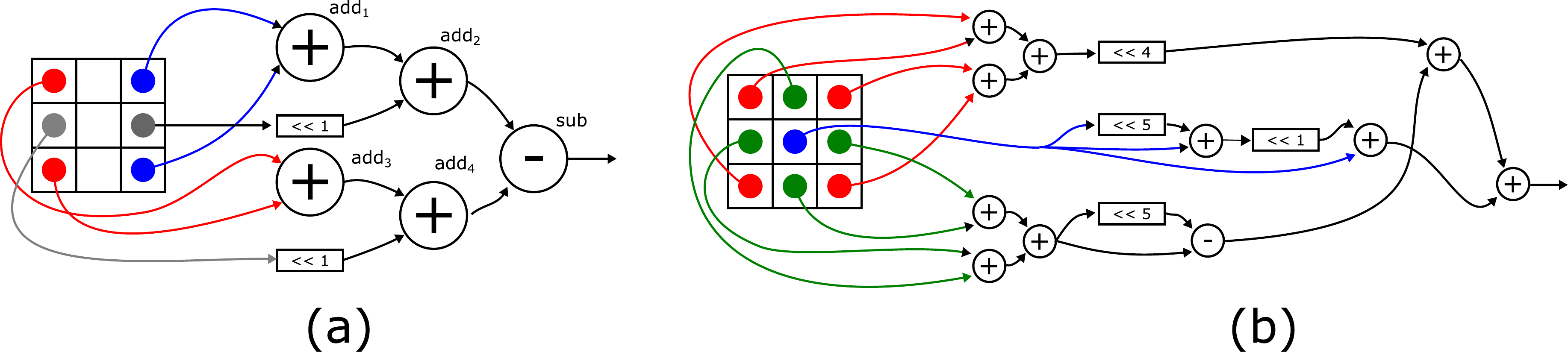}\vspace{-1em}
    \caption{Architecture of (a) Sobel edge detector; (b) fixed Gaussian filter}\vspace{-1.0em}
    \label{fig:sobel}
\end{figure}

We implemented \textit{Sobel edge detector} for detecting vertical edges. 
Its structure is shown in Figure \ref{fig:sobel}a. It consists of four adders, one subtractor and two shifts. 
For QoR analysis, 24 images from the benchmark data set were employed.

The filter kernel for the \textit{Fixed GF} was generated using the following parameters:  $w=3,\sigma=2$. 
Since the coefficients are constant, multiplierless constant multipliers (MCMs) can be employed. The architecture of this filter is shown in Figure~\ref{fig:sobel}b. The filter thus consists of adders, subtractors and shifts only. The optimum MCMs were obtained using \textit{SPIRAL tool}~\cite{spiral:2018}. For QoR analysis, 24 images from the benchmark data set were employed. 


Contrasted to the fixed GF, the generic GF is, in fact, a common convolution filter with variable kernel coefficients. The hardware model consists of nine 8-bit multipliers whose results are summed. To evaluate QoR, we created a C++ model which considers $50$ different Gaussian kernels generated for $w=3$ and $\sigma$ ranging from $0.3$ to $0.8$. 
Four images were selected from the bechmark dataset. In total, $200$ different simulations have been performed during QoR and the average SSIM was used as the quality indicator.


\newcommand{\csk}{\hspace{3mm}}


\section{Results}
The results are divided into two parts. Firstly, a detailed analysis of the results for the Sobel ED is provided to illustrate the principle of the proposed methodology. In the second part, only the final results are discussed due to the complexity of these problems and a limited space.

\subsection{Sobel edge detector}
\subsubsection{Library pre-processing}
\begin{figure}[ht]
    \centering
    \vspace{-1em}
    \includegraphics[width=\columnwidth]{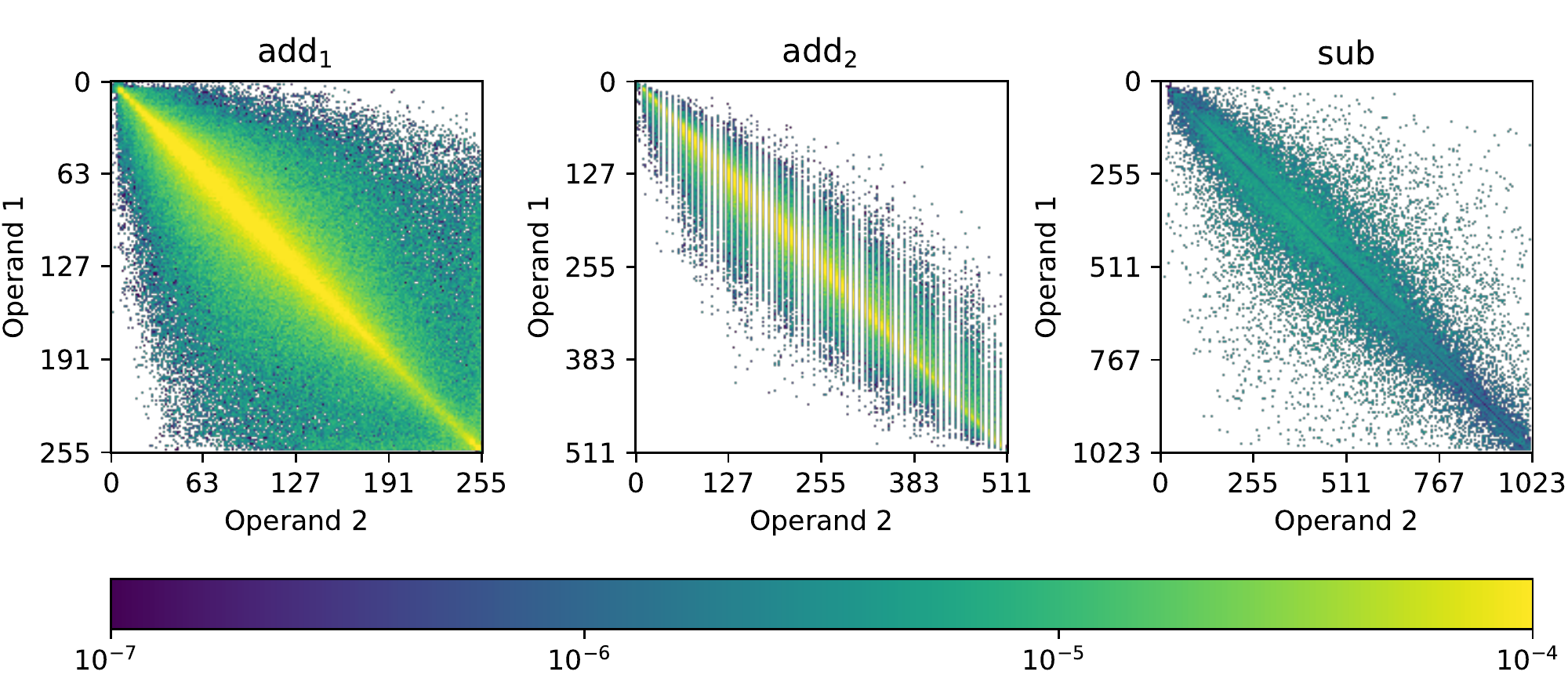}\vspace{-1em}
    \caption{Probability mass function of operations in the Sobel ED}
    \label{fig:profile}
    \vspace{-1em}
\end{figure}

To eliminate irrelevant circuits from the library, a score is calculated for each circuit in the library. Firstly, the target accelerator is profiled with a profiler which calculates the probability mass functions $D_k$ for all operations (Figure~\ref{fig:profile}). Note that $add_3$ (resp. $add_4$) has almost identical PMF with $add_1$ (resp. $add_2$). Figure~\ref{fig:profile} shows that operand values (neighbour pixels) are typically very close. In the plot dealing with $D_{add_2}$ one can see regular white stripes caused by shifting of the second operand.

Using the obtained probabilities, we calculated $WMED_k$ for all approximate circuits implementing $k$-th operation. Then we executed  a component filtering process guided by  $area$ and $WMED_k$ parameters of the isolated circuits and kept only Pareto optimal implementations. 
At the end of this process, the number of circuits in reduced libraries is $|RL_{add_1}|=35, |RL_{add_2}|=32, |RL_{add_3}|=37, |RL_{add_4}|=33,$ and $|RL_{sub}|=36$.

\subsubsection{Model construction}
The next step in the methodology is to construct models estimating SSIM and hardware parameters using parameters of the circuits belonging to one selected configuration. We used $WMED$ of all employed circuits as the input vector for the QoR model. For the hardware model we used $power, area$ and $delay$ of all circuits as the input vector. Several learning engines were compared to identify the most suitable one for our methodology (1500 configurations for learning and 1500 configurations for testing were randomly generated using the reduced libraries).

The considered learning engines were the regression algorithms from \textit{scikit-learn} tool for Python. Additionally, we constructed na\"ive models for area ($M_a(C) = \sum_{\forall c \in C}{area(c)}$) and for SSIM ($M_{SSIM}(C) = -\sum_{\forall c \in C}{WMED_k(c)}$) to test if SSIM correlates with the cumulative arithmetic error and if the area correlates with the sum of areas of all employed circuits. These simple models were also considered in our comparisons.

\begin{table}[h]
    \vspace{-0.5em}
    \caption{The fidelity of models for Sobel edge detector constructed by different learning engines. }\label{tab:regression}\vspace{-1em}
    \input{tab_regression.tex}
    \vspace{-1em}
\end{table}

\begin{figure}[b]
    \centering\vspace{-1em}
    \includegraphics[width=\columnwidth]{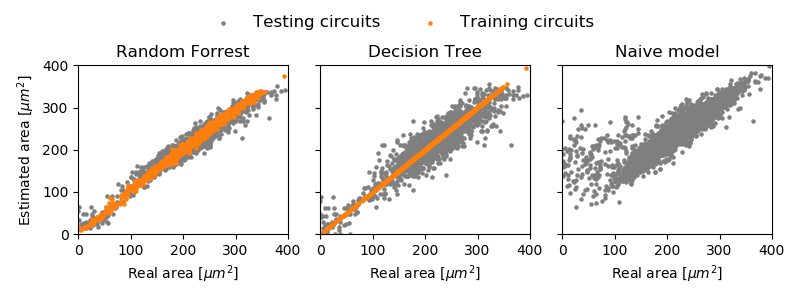}\vspace{-2em}
    \caption{Correlation of estimated area and real area obtained by synthesis tool for the selected learning engines used in Sobel ED experiment.}
    \label{fig:regression}\vspace{-1.5em}
\end{figure}

Table~\ref{tab:regression} shows the fidelities for all constructed models when evaluated on the training and testing data sets. The best result for the testing data sets are provided by random forest consisting of 100 different trees. The correlation between estimated and real area is shown in Figure~\ref{fig:regression}. 
The na\"ive models exhibit unsatisfactory results especially for small resulting approximate accelerators. When we analyze some of these cases in detail we observe that the inaccuracy was typically caused  by the last operation in the application (i.e., \textit{sub}). As this operation shows a big error, it is significantly simplified by the synthesis tool and as a consequence of that many other circuits are removed from the circuit because their outputs are no longer connected to any component. Hence the real area of these circuits was significantly smaller than the area calculated using  the library. Due to this elimination, machine learning methods based on conditional structures (e.g., trees) exhibit better performance than methods primarily utilizing algebraic approaches (e.g., MLP NN).

We tried to understand the impact of input parameters on the model quality. Including different error metrics such as the error variance did not improve the fidelity of QoR models. In contrast, omitting of power and delay in hardware modeling led to 2\% lower fidelities of these models in average.

\subsubsection{Model-based design space exploration}
In this part, the quality of proposed heuristic algorithm that we used for Pareto frontier construction is evaluated. Because of a low number of operations in Sobel ED, we are able to evaluate all possible configurations derivable from the reduced libraries $RL_k$ (i.e., $4.92 \cdot 10^7$ configurations in total). Note that the limit for stagnation detection was set to $50$ iterations in Alg. 1.

\begin{table}[bht]
    \centering
    \caption{Distances of the configurations identified by the proposed algorithm and random search from the optimal Pareto front. The lower value the better.}
    \label{tab:distances}  \vspace{-1em}  \small
    {\setlength\tabcolsep{2pt}\begin{tabular}{l c c  |c c | c c }\toprule
\multirow{2}{*}{\bf Algorithm}  & \multirow{2}{*}{\bf \#eval}   & \multirow{2}{*}{\bf \#Pareto} 	&	\multicolumn{2}{c|}{\bf To optimal}	&	\multicolumn{2}{c}{\bf From optimal} \\
 & 	& &	\it avg	&	\it max	&	\it avg	&	\it max \\\midrule
Optimal Pareto & $5\cdot10^7$ & 335 & --- & --- & --- & --- \\\midrule
    & $10^3$ & 71 &	0.02538	&	0.07554	&	0.03318	&	0.08650\\

Proposed  & $10^4$ & 177	&	0.00253	&	0.01328	&	0.00341	&	0.01690\\
    & $10^5$ & \bf 324	& \bf 0.00001	& \bf 0.00095	&	\bf 0.00009	& \bf	0.00657\\\midrule

    & $10^3$ & 37	&	0.05276	&	0.10615	&	0.05616	&	0.11307\\
Random sampling  & $10^4$ & 61	&	0.02631	&	0.08981	&	0.02875	&	0.07215\\
    & $10^5$ & 82	&	0.01172	&	0.03770	&	0.01353	&	0.03820\\\bottomrule
    \end{tabular}}
\end{table}

Pareto fronts created by means of the proposed algorithm were compared with Pareto fronts constructed using the random sampling (RS) algorithm and the optimal Pareto fronts.
The results are summarized in Table~\ref{tab:distances}.  We can see that the proposed algorithm with $10^5$ evaluations allows us to get almost the same number of Pareto configurations as the optimal Pareto front contains. To show that obtained configurations $S$ are very close to the optimal configurations $P$, the distances of obtained configurations to the nearest optimal configuration $(\forall s \in S: \min_{\forall p \in P}|s - p|)$ and the distances from the optimal configuration to the nearest obtained configurations $(\forall p \in P: \min_{\forall s \in S}|s - p|)$ are analyzed. Both algorithms provided configurations that are typically close to the optimal one, but RS missed a lot of important configurations. Note that the distance is calculated from estimated QoR and HW parameters normalized to range \textit{<0,1>}. 

\subsection{Gaussian filters}
The methodology was also applied to obtain approximate implementations of two versions of Gaussian image filter (fixed GF and generic GF). After profiling this accelerator and reducing the library of approximate circuits accordingly, random forest-based models of QoR and hardware parameters were created using 4000 training and 1000 testing randomly generated configurations. In the case of fixed GF, the fidelity of the area estimation model is 87\% for hardware parameters and 92\% for QoR. The fidelity of both models of generic GF is 89\%. If the synthesis and simulations run in parallel, the detailed analysis of one configuration takes $10$~s on average and the model-based estimation of one configuration takes $0.01$~s on average.

The Pareto construction algorithm evaluated $10^6$ candidate solutions. On average,  $39$ iterations were undertaken to find a new candidate suitable for the Pareto front.

\begin{table}[tb!]
    \centering
    \caption{Size of the design space after performing particular steps of the proposed methodology}
    \label{tab:space}\vspace{-1em}\small
    {\setlength\tabcolsep{2pt}\begin{tabular}{l | c c c c}
    \toprule \multirow{2}{*}{\bf Application} & \multicolumn{4}{c}{\bf \# configurations} \\
                            & \it  all possible & \it lib. pre-processing & \it pseudo Pareto  & \it final Pareto \\\midrule
Sobel ED    & $1.96 \cdot 10^{15}$ & $4.92 \cdot 10^{7}$  & $335$ & $62$ \\
Fixed GF & $7.35 \cdot 10^{34}$ & $1.73 \cdot 10^{16}$ & $1166$ & $132$ \\
Generic GF & $7.15 \cdot 10^{63}$ & $3.75 \cdot 10^{23}$ & $946$ & $102$ \\\bottomrule
    \end{tabular}\vspace{-1em}
    }
\end{table}

Table~\ref{tab:space} shows the size of the design space after performing particular steps of the proposed methodology. For example, there are $7.15\cdot10^{63}$ configurations in the generic GF design space. The elimination of irrelevant circuits in the library reduced the number of configurations to $3.75\cdot 10^{23}$. The number of configurations is enormous because it would take $10^{17}$ years to analyze them. In contrast, the construction of 4000 random solutions for training of the models takes approximately 11 hours, $10^6$ iterations of the proposed Pareto construction algorithm employing the models takes $3$ hours and the remaining $1000$ configurations are analyzed in $3$ hours. Finally, approximately $100$ configurations that are Pareto optimal in terms of area, SSIM and energy are selected. In total, the proposed approach takes $17$ hours on a common desktop. Hypothetically, if we would use the analysis instead of the estimation model in the Pareto front construction, the analysis of $10^6$ configurations would take $115$ days. 



\begin{figure}[b!]
    \centering\vspace{-10pt}
    \includegraphics[width=\columnwidth]{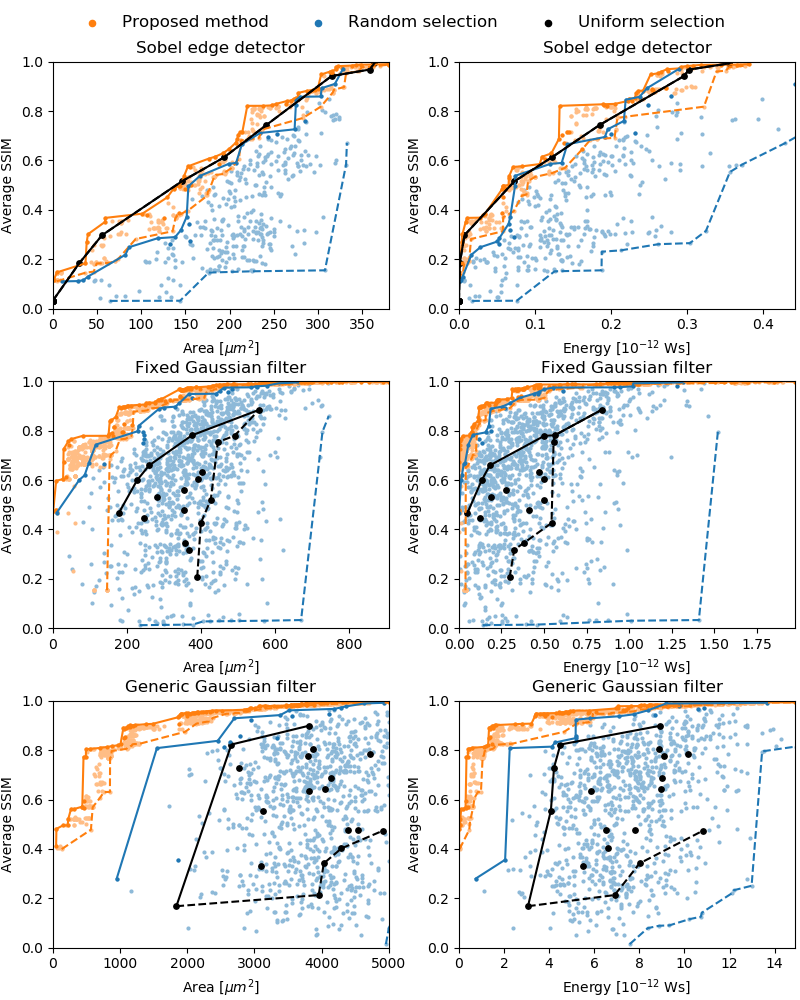} \vspace{-2.5em}
    \caption{Pareto fronts showing best tradeoffs between SSIM, area and energy obtained using three methods (orange -- the proposed method; blue -- RS; black -- uniform selection) for three approximate accelerators. }
    \label{fig:quality}
\end{figure}

Figure~\ref{fig:quality} compares resulting Pareto fronts obtained using the proposed methodology (orange line), the RS-based Pareto front construction algorithm (blue line) and the uniform selection approach (black line). The uniform selection approach is a manual selection method which one would probably take if no automated design methodology is available. In this method, particular approximate circuits are deterministically selected to exhibit the same error WMED (relatively to the output range). Figure~\ref{fig:quality} shows that this method provides relevant results only for accelerators containing a few operations. The randomly generated configurations (blue points) were obtained from a 3 hour run of the random configuration generation-and-evaluation procedure. They are included to these plots in order to emphasize high quality solutions obtained by the proposed method. 

\section{Conclusions}
We developed an automatic design space exploration and circuit approximation methodology which replaces operations in an original accelerator by their approximate versions taken from a library of approximate circuits. In order to accelerate the approximation process, QoR and hardware parameters are estimated using computational models created by means of machine learning methods. On three case studies we have shown that the proposed methodology provides approximate accelerators showing high-quality tradeoffs between QoR and hardware parameters. Our methodology paves a way towards a fully automated approximation of complex accelerators that are composed of approximate operations whose error models are in principle unknown.

\paragraph{Acknowledgments}
\footnotesize{
This work was supported by Czech Science Foundation project 19-10137S and by the Ministry of Education of Youth and Physical Training from the Operational Program Research, Development and Education project International Researcher Mobility of the Brno University
of Technology --- CZ.02.2.69/0.0/0.0/ 16\_027/0008371}

%% file: algorithmv2.tex
\begin{flushleft}
        \textbf{INPUT:} $RL$  -- set of libraries, $RL=\{RL_1,RL_2,\cdots,RL_n\}$, \\
        \hspace{8mm} $M_{HW}$ -- HW costs model, $M_{QoR}$ -- quality model  \\
        \textbf{OUTPUT:} Pareto set $P \subseteq RL_1 \times RL_2 \times \cdots \times RL_n$ 
\end{flushleft}
\begin{algorithmic}

\Function{HeuristicParetoConstruction}{$RL$, $M_{QoR}$, $M_C$}
\State $Parent \gets \Call{PickRandomlyFrom}{ RL_1 \times RL_2 \times \cdots \times RL_n}$
\State $P \gets \emptyset$
\While{$\neg TerminationCondition$}
    \State $C \gets \Call{GetNeighbour}{Parent}$
    \State $e_{QoR} \gets M_{QoR}(C)$ \Comment{\footnotesize Estimate the quality of C}
    \State $e_{HW} \gets M_{HW}(C)$ \Comment{\footnotesize Estimate the HW costs of C}
    \If {\Call{ParetoInsert}{$P, (e_{QoR}, e_{HW}), C$}} 
        \State $Parent \gets C$
    \ElsIf{\it StagnationDetected} \Comment{\footnotesize Parent not changed in last $k$ iterations}
        \State $Parent \gets \Call{PickRandomlyFrom}{P}$
    \EndIf

\EndWhile\label{euclidendwhile}
\State \Return{P}
\EndFunction
\end{algorithmic}

%% file: tab_regression.tex
\small\begin{tabular}{l|c c|c c}\toprule
\multirow{2}{*}{\bf Learning algorithm}	 & 	\multicolumn{2}{c|}{\bf SSIM}	 		 & 	\multicolumn{2}{c}{\bf Area} \\
	 & \it	Train	 & \it	Test	 & 	\it Train	 & \it 	Test\\\midrule
Random Forest	 & 	99\%	 & 	\bf 96\%	 & 	97\%	 & \bf	92\%\\
Decision Tree	 & \bf	100\%	 & 	95\%	 & \bf	100\%	 & 	86\%\\
K-Neighbors	 & 	94\%	 & 	94\%	 & 	91\%	 & 	89\%\\
Bayesian Ridge	 & 	90\%	 & 	90\%	 & 	91\%	 & 	91\%\\
Partial least squares	 & 	90\%	 & 	90\%	 & 	91\%	 & 	90\%\\
Lasso	 & 	90\%	 & 	90\%	 & 	91\%	 & 	90\%\\
Na\"ive model	 & ---	 & 	90\%	 & 	---	 & 	88\%\\
Ada Boost	 & 	90\%	 & 	90\%	 & 	90\%	 & 	88\%\\
Least-angle	 & 	90\%	 & 	90\%	 & 	71\%	 & 	72\%\\
Gradient Boosting	 & 	89\%	 & 	89\%	 & 	92\%	 & 	91\%\\
MLP neural network	 & 	86\%	 & 	83\%	 & 	92\%	 & 	91\%\\
Gaussian process	 & 	\bf 100\%	 & 	71\%	 & \bf	100\%	 & 	55\%\\
Kernel ridge	 & 	41\%	 & 	42\%	 & 	90\%	 & 	90\%\\
Stochastic Gradient Descent	 & 	24\%	 & 	25\%	 & 	75\%	 & 	74\%\\
\bottomrule
\end{tabular}